\documentclass[12pt]{article}
\usepackage{graphicx}
\usepackage{amsmath}

\bigskip
\input{tcilatex}

\begin{document}

\title{Integral Geometry on the Lobachevsky Plane and the Conformal
Wess-Zumino-Witten Model of \ Strings on an ADS$_{3}$ Background}
\author{Bogdan G. Dimitrov \\
{\small \textit{Bogoliubov Laboratory of Theoretical Physics}}\\
[-1.mm] {\small \textit{Joint Institute for Nuclear Research, Dubna 141 980,
Russia}}\\
[-1,mm] {\small \textit{email: bogdan@thsun1.jinr.ru}}}
\date{}
\maketitle

\begin{abstract}
The main purpose of \ the \ report is to provide some argumentation that
three seemingly distinct approaches of 1. Giveon, Kutasov and Seiberg
(hep-th/9806194); 2. Hemming, Keski-Vakkuri (hep-th/0110252); Maldacena,
Ooguri (hep-th/0001053) and 3. I. Bars (hep-th/9503205) can be investigated by
applying the mathematical methods of integral geometry on the
Lobachevsky plane, developed previously by Gel'fand, Graev and Vilenkin. All
these methods can be used for finding the transformations, leaving the
Kac-Moody and Virasoro algebras invariant. The near-distance limit of the
Conformal Field Theory of \ the SL(2, R)\ WZW model of strings on an ADS$%
_{3} $ background can also be interpreted in terms of the Lobachevsky
Geometry : the non - euclidean distance is conserved and the Lobachevsky
formulae for the angle of parallelism is recovered. Some preliminary
technique from integral geometry for inverting the modified integral
representation for the Kac- Moody algebra has been demonstrated.

\end{abstract}

\section{Introduction and Statement of the Problem}

The present paper has the purpose to illustrate the importance of the ideas
and constructions of the Non-Euclidean (Lobachevsky) Geometry, which can be
applied even today for solving some conceptually important problems in
\textbf{string theory on an }$\ $\textbf{(Anti - DeSitter) }$ADS_{3}$\textbf{%
\bigskip\ background. }Presently one of the most widely discussed topics is
the $ADS/CFT$ correspondence, where the $3D$\textbf{\ ''Anti-DeSitter''}
space is in fact the $3D$ \textbf{Lobachevsky space of constant negative
curvature}, proposed first by Lobachevsky, Boyai (subsequently by Gauss,
Beltrami and many others) yet in the 18th century. In this aspect one should
mention also the lectures of \ N.A. Chernikov \cite{1}, which review and
clarify such fundamental notions as the horosphere, the horocycles, the
Poincare model and etc., constituting the basic necessary \ \ knowledge for
application of the integro-geometric approach of Gel'fand, Graev and
Vilenkin \cite{2, 3}.

In the last several years there has been a considerable progress in
constructing \textbf{Conformal Field Theories} \textbf{(CFT}) on an \textbf{%
Riemann surface} \cite{4, 5}. One of the main successes of this theory \ is
that on the base of the complex automorphisms of the Riemann sphere
(isomorphic to the group $SL(2,C)/Z_{2}$ and represented as a composition of
\textbf{translation, the scaling transformation and the special conformal
transformation) } and by means of the duality theorem for the (specially
constructed) \textbf{vertex operator }$V$, it turns out to be possible to
derive the \textbf{Virasoro algebra }
\begin{equation}
\left[ L_{m},L_{n}\right] =(m-n)L_{m+n}+\frac{c}{12}m(m^{2}-1)\delta _{m+n,0}
\tag{1}
\end{equation}
\textbf{without the use of the OPE\ relations }
\begin{equation}
T(z)T(v)=\frac{c}{2(z-v)^{4}}+\frac{2}{(z-v)^{2}}T(v)+\frac{1}{z-v}%
T^{^{\prime }}(v)+......\text{ \ \ \ \ ,}  \tag{2}
\end{equation}
\begin{equation}
T(z)J(v)=\frac{J^{a}(v)}{(z-v)^{2}}+\frac{1}{(z-v)}J^{a^{\prime }}(v)+......%
\text{ \ \ \ \ ,}  \tag{3}
\end{equation}
\begin{equation}
J^{a}(z)J^{b}(v)=\frac{k\delta ^{ab}}{(z-v)^{2}}+\frac{f^{abc}}{(z-v)}%
J^{c}(v)+......\text{ \ \ \ \ }  \tag{4}
\end{equation}
for the stress tensor $T(z)$ and for the conformal current $J^{a}(z)$ -
unlike the approach in the first papers on CFT \cite{6, 7, 8}. Three major
developmens in the contemporary theory will be mentioned, which may seem to
be distinct, but in fact the purpose here will be to show that an unifying
mathematical approach can be proposed \ in respect to all of them.

\textbf{The first development} is related to the problem about the \textbf{%
spectral flow} \cite{9, 10, 11} ,\textbf{\ }which generates new classical
solutions in the $WZW$ model after applying the operation
\begin{equation}
g_{+}=exp\left[ \frac{i}{2}\omega _{R}x^{+}\sigma _{2}\right] \widetilde{g}%
_{+}(x^{+})\text{ \ \ \ \ ; \ \ \ \ \ \ \ }g_{-}=\widetilde{g}_{-}(x^{-})exp%
\left[ \frac{i}{2}\omega _{L}x^{-}\sigma _{2}\right] \text{ \ ,}  \tag{5}
\end{equation}
where the group element $g$ factorizes into the product of the right and
left group elements $g_{+}$ and $g_{-}$ , $x^{+}=\tau +\sigma $ $\ $and $%
x^{-}=\tau -\sigma $ \ are the (right and left) coordinates on the
worldsheet and \ $\omega _{L}$, $\omega _{R}$. are the eigenvalues of the
left and right conformal currents. For the case of the BTZ black hole \cite
{11}, the spectral flow \textbf{defines a mapping between the string
worldsheet coordinates }$\tau $\textbf{,}$\sigma $\textbf{\ and the global
spacetime coordinates }$r,\widehat{t},\widehat{\Phi }$ \
\begin{equation}
\widehat{t}\rightarrow t+\frac{1}{2}\left[ (\omega _{+}+\omega _{-})\tau
+(\omega _{+}-\omega _{-})\sigma \right]  \tag{6}
\end{equation}
\begin{equation}
\widehat{\Phi }\rightarrow \Phi +\frac{1}{2}[(\omega _{+}+\omega _{-})\sigma
+(\omega _{+}-\omega _{-})\tau ]\text{ \ \ \ \ \ \ ,}  \tag{7}
\end{equation}
under which the components of the conformal currents and of the Virasoro
generators should transform as
\begin{equation}
\widetilde{J}_{R}^{2}=J_{R}^{2}+\frac{k}{2}w_{+}\text{ \ \ \ \ ; \ \ \ \ \ }%
\widetilde{J}_{R}^{\pm }=J_{R}^{\pm }exp(\mp w_{+}x^{+})  \tag{8}
\end{equation}
\begin{equation}
\widetilde{L}_{n}=L_{n}+w_{+}J_{n}^{2}+\frac{k}{4}w_{+}^{2}\delta _{n,0}%
\text{ \ \ \ \ \ \ ,}  \tag{9}
\end{equation}
so that the invariance of the Kac - Moody and Virasoro algebras is ensured.

\textbf{The second development} is proposed in a paper of Giveon, Kutasov
and Seiberg \cite{12}. Let $Q_{0}^{a}$ are the \textbf{spacetime conserved
charges}
\begin{equation}
Q_{0}^{a}\equiv \oint dzJ^{a}(z)\text{ \ \ \ ,}  \tag{10}
\end{equation}
which evidently satisfy the closed Lie algebra $\left[ Q_{0}^{a},Q_{0}^{b}%
\right] =if_{c}^{ab}Q_{0}^{c}$ and the conformal currents $J^{a}(z)$ ($%
a=3,+,-$ for the $SL(2,R)$ case) satisfy the OPE relations (4), where $k$ is
the level of the string worldsheet algebra. In order to \textbf{extend the
zero - mode symmetry of the space time charges }$Q_{0}^{a}$\textbf{\ to an
infinite symmetry of the charges }$Q_{n}^{a}$, satisfying the affine Lie
algebra
\begin{equation}
\left[ Q_{n}^{a},Q_{m}^{b}\right] =if_{c}^{ab}Q_{n+m}^{c}+\frac{\widetilde{k}%
}{2}n\delta ^{ab}\delta _{n+m,0}\text{ \ \ ,}  \tag{11}
\end{equation}
the authors propose to define $Q_{n}^{a}$ by multiplying the integrand $%
J^{a}(z)$ in (10) with a holomorphic function $\gamma ^{n}(z)$, i. e $%
Q_{n}^{a}\equiv \oint dzJ^{a}(z)\gamma ^{n}(z)$. It is important that the
function $\gamma ^{n}(z)$ comes from the ''microscopic'' definition of a
\textbf{vertex operator \cite{13} in string theory} $V_{jm\overline{m}%
}=\gamma ^{j+m}\gamma ^{-j+\overline{m}}exp(\frac{2j}{\alpha }\Phi )$, this
time for values of the quantum numbers $j=0,\overline{m}=0,m=n$. Then the
''modified'' charge operators $Q_{n}^{a}$ will satisfy the affine Lie
algebra (11) only if the \textbf{level }$\widetilde{k}$\textbf{\ of the
affine Lie algebra in spacetime} is $p-$times the \textbf{level }$k$\textbf{%
\ (see \ eq. 4) of \ the \ string worldsheet algebra}, i.e. $\widetilde{k}%
=pk $. The number $p\equiv \oint dz\frac{\partial _{z}\gamma }{\gamma }$
''counts'' the number of times the string worldsheet wraps around the
angular coordinate $\Theta $.

Now what is the conclusion from all these reasonings? The function $\gamma
^{n}(z)$ in the ''microscopic'' definition of the vertex operator also
enters in the expression for the metric
\begin{equation}
ds^{2}=L^{2}(d\Phi ^{2}+e^{2\Phi }d\gamma d\overline{\gamma })\text{ \ \ \ ,}
\tag{12}
\end{equation}
obtained after a given parametrization of the global $ADS3$ (Lobachevsky)
coordinates $(X_{0},X_{1},X_{2},X_{3})$, defined on the hyperboloid $%
X_{0}^{2}-X_{1}^{2}-X_{2}^{2}-X_{3}^{2}=L^{2}$. Furthermore, the conformal
currents in the $WZW$ model
\begin{equation}
J_{L}^{a}=kTr(T^{a}g^{-1}\partial _{\overline{z}}g)\text{ \ \ \ \ \ \ \ \ \
\ \ \ \ \ \ \ \ \ \ \ }J_{R}^{a}=kTr(T^{a}\partial _{z}g.g^{-1})\text{ \ \ \
,}  \tag{13}
\end{equation}
in which the group element $g$ is parametrized by means also of \ the global
coordinates $(X_{0},X_{1},X_{2},X_{3})$ (see for some details \cite{14}),
are identified with the currents in the OPE relations and the Kac - Moody
and Virasoro algebras in the $WZW$ model. \textbf{Therefore, we come to the
following important problem:\ How are the global symmetries of the }$ADS3$%
\textbf{\ (Lobachevsky)\ spacetime related to the transformations, which
leave the Kac - Moody and Virasoro algebras invariant?}

To confirm the importance of the stated problem, \ let us mention briefly
the \textbf{third development} in the string theory on an $ADS3$ spacetime
in some of the papers of \ I. Bars \cite{15}. He assumes that there is
\textbf{an additional zero mode} (proportional to $lnz$ -pieces) that is
present in the local conserved $SL(2,R)$ currents of the $WZW$ model.
Similarly to the previous developments, in \cite{15} and some other papers,
the transformations of the conformal currents (and of the stress tensor),
leaving the Kac-Moody and Virasoro algebras invariant are also found.

\section{\protect\bigskip Basic Assumptions, Objectives of the Present
Research and Results, Related to Lobachevsky Geometry}

\bigskip The present paper and the performed research as a whole have the
following objectives:

1. To clarify the non-euclidean geometrical meaning of the found in \cite{14}
algebraic relation in \ $CP3$ from the OPE relations with the conformal
current (only). A central result will be that the \textit{non-euclidean
distance }$\rho $, defined by means of the \textbf{anharmonic relation}
\begin{equation}
\rho =rln\mid (w\text{ }z\text{ }v\text{ }t)\mid =r\text{ }ln\mid \frac{%
(w-v)(z-t)}{(z-v)(w-t)}\mid  \tag{19}
\end{equation}

for four points $w,$ $z,$ $v$ and $t$ on the Lobachevsky (hyperbolic) plane (%
$r$ is a scale factor), is a \textbf{constant} (complex number). Since no
OPE relations with the stress tensor have been used and therefore there is
\textbf{no translation, this might be interpreted as being physically
consistent with the initial assumption about rotations (only). }

2. After an identification of the conformal currents in the OPE relations
with the conformal current in the $WZW$, to investigate the $SL(2,R)$ OPE
relations. As \textbf{one} \textbf{condition} (several others have been
found, but represented by much more complicated expressions) for \textbf{%
consistency of the OPE relations} and \ after some redefinitions of the
parametrization variables $(\Psi ,\Theta ,\Phi )$, the well-known \textbf{%
Lobachevsky formulae \cite{1, 16} for the angle of parallelism} will be
recovered:
\begin{equation}
\Pi (\rho )=2arctg(exp(-\frac{\rho }{R}))\text{ \ \ \ \ ,}  \tag{20}
\end{equation}
where $\rho $ is the noneuclidean distance (19) \ for \ the \ worldsheet
manifold and the number $R$ is the so called ''curvature radius'' of the
Lobachevsky plane - a measure for the ''noneuclideancy'' of the Lobachevsky
geometry.

Therefore, the applicability of \ Lobachevsky geometry for treating the $%
SL(2,R)$ OPE relations has been confirmed \textbf{twice: }before and after
the identification of the conformal currents. This can hardly be a simple
coincidence of calculations.

In both cases only the pole terms in the OPE relations have been taken into
account and all the s. c. ''regular'' terms have been discarded.
Consequently, this is the ''near - distance'' approximation limit of the
points on the complex plane, which is in accord with taking into account
only the first two terms in the variation $\delta _{\varepsilon
}A_{j}(z)=\sum\limits_{k=0}^{\nu _{j}}B_{j}^{(k-1)}(z)\frac{d^{k}}{dz^{k}}%
\varepsilon (z)$ ($\nu _{j}$- integer numbers) of the local field $A_{j}(z)$
\cite{5}. This comes from the requirements $\Delta _{j}\geq 0$ ($\Delta _{j}$%
- conformal dimension of the local fields $A_{j}(z)$; $\Delta _{j,(k-1)}$ -
of the fields $B_{j}^{(k-1)}(z)$) $\Delta _{j,(k-1)}=\Delta _{j}+1-k$ and $%
\Delta _{j,(k-1)}$ $=0$ (the lowest possible value), which guarentee that
any correlators in the CFT\ would be convergent with the increase of
distance (see \cite{6}).

3. In order to incorporate the symmetries of the global $ADS$ coordinates in
the problem about the transformations, leaving the \ Kac- Moody and Virasoro
algebras invariant, and with the purpose to unite the three distinct
approaches, discussed in the Introduction, in \cite{14} it was proposed to
\textbf{modify the integral definition} of the Virasoro and of the conformal
current generators (here only the formulae for the Virasoro generators will
be written)
\begin{equation}
\widehat{L}_{n}(\eta )A_{j}(z,\overline{z})=\oint\limits_{C}\left[
\oint\limits_{C_{1}}T(\zeta )(\zeta -z)^{n+1}\delta (\left[ \widetilde{X}%
,\eta \right] +1)A_{j}(z,\overline{z})d\widetilde{X\text{ }}\right] d\zeta
\text{ \ \ \ \ ,}  \tag{21}
\end{equation}
where (in the spirit of \ the integro - geometric approach of Gel'fand,
Graev and Vilenkin \cite{2, 3}) an additional integration is introduced on
the $ADS$ hyperboloid coordinates $\widetilde{X\text{ }}$ ($d\widetilde{X%
\text{ }}\equiv d\widetilde{X\text{ }}^{0}d\widetilde{X\text{ }}^{1}d%
\widetilde{X\text{ }}^{2}d\widetilde{X\text{ }}^{3}$) and the presence of
the delta - function signifies that the integration is on the \textbf{%
horosphere }
\begin{equation}
\left[ \widetilde{X},\eta \right] +1=-X^{0}\eta ^{0}+X^{3}\eta
^{3}+X^{1}\eta ^{1}+X^{2}\eta ^{2}+1=0\text{ \ \ \ \ \ \ \ ,}  \tag{22}
\end{equation}
intersecting the cone $\left[ \eta ,\eta \right] =0$. \textbf{By definition,
the horosphere in Lobachevsky geometry is the surface, orthogonal to the set
of parallel lines, passing through one and the same point on the absolute
\cite{1, 16}, and the absolute by itself is simply the set of ''points'' (a
''point' may be a straight line, or a hyperplane) at infinity. The most
essential property of the horoshere is that on it the usual Euclidean
geometry is realized. }That is why the ''harmonic'' measure, defined by%
\textbf{\ }$\delta (\left[ \widetilde{X},\eta \right] +1)d\widetilde{X\text{
}}$ is invariant under rotations around the $\widetilde{X\text{ }}$-point.

The most important result in \cite{14} was that the Virasoro algebra \ $%
\left[ L_{n},J_{m}^{a}\right] =-mJ_{n+m}^{a}$ in terms of the newly defined
operators can be written as
\begin{equation}
\oint\limits_{(\zeta _{1},X_{1})}F(\zeta _{1},\zeta _{2},z)\delta (\left[
X_{1},\eta _{1}\right] +1)d\zeta _{1}d\widetilde{X}=-mJ^{a}(\zeta
_{2})(\zeta _{2}-z)^{m+n}\text{ \ \ \ \ ,}  \tag{23}
\end{equation}
where $F(\zeta _{1},\zeta _{2},z)\equiv T(\zeta _{1})J^{a}(\zeta _{2})\left[
(\zeta _{1}-z)^{n+1}(\zeta _{2}-z)^{m}t_{j}^{m}-(\zeta _{1}-z)^{m}(\zeta
_{2}-z)^{n+1}t_{j}^{n}\right] $. Now one way of realizing the large - scale
symmetries is when the integration over the $d\widetilde{X}\equiv d%
\widetilde{X}^{0}d\widetilde{X}^{1}d\widetilde{X}^{2}d\widetilde{X}^{3}$
variables is performed, making the variable change $q=\left[ \widetilde{X}%
,\eta \right] +1$ , integrating over $\int \frac{\partial X^{0}}{\partial q}%
\delta (q)dqd\widetilde{X}^{1}d\widetilde{X}^{2}d\widetilde{X}^{3}$ and
lastly, taking into account the hyperboloid equation $-\left( \widetilde{X}%
^{0}\right) ^{2}-\left( \widetilde{X}^{1}\right) ^{2}+\left( \widetilde{X}%
^{2}\right) ^{2}+\left( \widetilde{X}^{3}\right) ^{2}=1$. This trivial
calculation shall not be performed here. The details, including an extensive
review of the approaches in integral geometry and of the basic notions in
Lobachevsky geometry, will be given in \cite{17}.

Of course, if one knows the algebraic surface $p(\zeta _{1},\zeta _{2},z)$,
on which these three complex variables are defined, then in the integral
(23) one may introduce an integration over the delta function $\delta
(p(\zeta _{1},\zeta _{2},z))$, which will allow to perform the integration
over the $\zeta _{1}$- variable in (23). Unfortunately, as we shall later
show (and was shown in \cite{14}), such an relation will be obtained from
the conformal OPE relations with the conformal current \textbf{only, so one
does not have the right to apply it to the }Virasoro algebra \ $\left[
L_{n},J_{m}^{a}\right] =-mJ_{n+m}^{a}$, where the Virasoro generator $L_{n}$
in (21) is related also to the stress tensor $T(\zeta )$. In principle, the
problem about deriving an algebraic relation from all the OPE relations
remains unsolved, and it requires significant efforts, since, as it is well
known, the transformation properties of the stress tensor are nontrivial -
it is transformed by means of \ the s. c. \textbf{Schwarzian}. Still,
another approach from integral geometry - the \textbf{Radon's transformation
in complex space} can be used, and its inverse one is well-known \cite{2}.

Further, it shall be proposed to take advantage of \ knowing the algebraic
relation
\begin{equation}
p(\zeta _{1},\zeta _{2},z)\equiv \frac{1}{(\zeta _{1}-z)^{2}}+\frac{1}{%
(\zeta _{2}-z)^{2}}=0\text{ \ \ \ \ }  \tag{24}
\end{equation}

for the complex points on the worldsheet, on which the integral operators in
the \textbf{Kac - Moody algebra}
\begin{equation}
\left[ J_{n}^{a},J_{m}^{b}\right] =if_{c}^{ab}J_{n+m}^{c}+\frac{k}{2}n\delta
^{ab}\delta _{n+m,0}  \tag{25}
\end{equation}
is defined. This time integrals of the type \cite{2, 3}
\begin{equation}
h(\eta )=\int f(X)\delta (\left[ X,\eta \right] -1)dX  \tag{26}
\end{equation}
are to be ''inverted''. The integral is defined on the upper-half $\ \left[
X,X\right] >0$, \ where all ''lines'' and ''hyperplanes'' remain inside the
absolute $\left[ X,X\right] =0.$ The inversion formulae for the the function
$f(X)$, taken at the point $X=a$, is \cite{2, 3}:
\begin{equation}
f(a)=-\frac{1}{8\pi ^{2}}\int \delta ^{^{\prime \prime }}(\left[ a,\eta %
\right] -1)h(\eta )d\eta \text{ \ \ \ \ .}  \tag{27}
\end{equation}
For the opposite case $\left[ X,X\right] <0$, when the so called ''imaginary
Lobachevsky space'' (on which the distance can be either a real number from $%
1$ to $\infty $, or an imaginary number), the inversion formulae is more
complicated, and shall not be considered here.

\section{Conservation of Non-Euclidean Distance from the OPE Relations for
the Conformal Current}

\bigskip The conservation of non-euclidean distance will be deduced from the
already found in \cite{14} algebraic relation (24) $\frac{1}{(z-v)^{2}}+%
\frac{1}{(w-v)^{2}}=0$ for the points $z,v$ and $w$. Let us remind briefly
that it was obtained from (4), written for $(A,B)=(3,+),$ $(3,-)$, $%
(+,-),(+,+),(-,-),(3,3)$, multiplying (4) consequently \ by $J^{C}(v)$ and
again applying the obtained OPE relations for the $SL(2,R)$ case.

Making use of the definition (19) for non-euclidean distance, it is easily
seen that the above algebraic relation can be rewritten as an \textbf{%
anharmonic relation for the four points }$w,z,v$\textbf{\ and }$t=\infty $:
\begin{equation}
\varepsilon i=\frac{w-v}{z-v}=(w\text{ }z\text{ }v\text{ }\infty )\text{ \ \
\ \ \ \ \ \ \ \ \ }(\varepsilon =+1\text{ \ }or\text{ }\varepsilon =-1)\text{
\ \ \ .}  \tag{28}
\end{equation}
The conservation of the non-euclidean distance immeadiately follows
\begin{equation}
\widetilde{\rho }=rln\mid (w\text{ }z\text{ }v\text{ }\infty )\mid =r\frac{%
i\pi }{2}\text{ \ if }\varepsilon =1\text{ and }=r\frac{3\pi i}{2}\text{ if }%
\varepsilon =-1  \tag{29}
\end{equation}
and therefore it determines an \textbf{equidistant surface in }$CP3$. From
hyperbolic geometry it is known that a \textbf{mapping of points is a Mobius
one if and only if the anharmonic relation is conserved.} But since Mobius
invariance is the basic assumption in the OPE relations, we receive a
consistent result, but this time from an algebraic point of view!

The point at infinity can be chosen also at $z=\infty $ or at $w=\infty $.
Then, writing down the algebraic relation for the points $v,w$ and $t$ and
combining the two relations, the conservation of the sum of two noneuclidean
distances can be obtained:
\begin{equation}
ln\mid (t\text{ }\infty \text{ }w\text{ }v)\mid +\text{ }ln\mid (t\text{ }z%
\text{ }v\text{ }\infty )\mid =0\text{ if }\varepsilon =1\text{ and \ }=r%
\text{ }exp.(\frac{i\pi }{2})\text{ \ if }\varepsilon =-1\text{ . }  \tag{30}
\end{equation}

\bigskip However, in terms of the real variables $%
x_{1},y_{1},x_{2},y_{2},x_{3},y_{3}$, where $z=x_{1}+iy_{1},$\ $%
v=x_{2}+iy_{2}$ \ \ and\ \ $w=x_{3}+iy_{3}$, the conservation of the
non-euclidean distance is not so obvious. After solving the equations for
the real and the imaginary parts of the algebraic relation, one gets
\begin{equation}
\frac{x_{1}-x_{2}}{x_{3}-x_{2}}.\frac{y_{1}-y_{2}}{y_{3}-y_{2}}=-1\text{ \ \
\ \ \ \ \ ,}  \tag{31}
\end{equation}
which can be represented also as
\begin{equation}
-1=(x_{1}\text{ }x_{3}\text{ }x_{2}\text{ }y_{2}).(y_{1}\text{ }y_{3}\text{ }%
y_{2}\text{ }x_{2}).F\text{ \ ; \ with }F\equiv \frac{%
(x_{1-}y_{2})(y_{1-}x_{2})}{(x_{3}-y_{2})(y_{3}-x_{2})}\text{ .}  \tag{32}
\end{equation}
Consequently, if $x_{2}\Longleftrightarrow y_{2}$ and $\widetilde{v}\equiv i%
\overline{v}$ satisfies the original algebraic relation (note that then $%
F=-1 $, i. e. $F$ is the original expression (31)), then a conservation of
the non - euclidean distance will also follow:
\begin{equation}
ln\mid (x_{1}\text{ }x_{3}\text{ }x_{2}\text{ }y_{2})\mid =ln\mid (y_{1}%
\text{ }y_{3}\text{ }x_{2}\text{ }y_{2})\mid \text{ \ .}  \tag{33}
\end{equation}
This relation in fact means that the non-euclidean distance is conserved in
respect to the symmetry changes $x_{1}\Longleftrightarrow y_{1}$ and $%
x_{3}\Longleftrightarrow y_{3}$. In the general case, however, \textbf{the
symmetries of the algebraic relation and of the expression for the
noneuclidean distance are different. }

\section{Recovering the Lobachevsky Formulae for the Angle of Parallelism
from the OPE\ Relations (4)}

\bigskip As already mentioned, the conformal currents in the $WZW$ model
(where the group element is parametrized by the $ADS$ global coordinates)
will be identified with the currents in the OPE relations. The following
additional and very simple assumptions will hold:

1. \textbf{The }$ADS$\textbf{\ global coordinates }$X^{\mu }$\textbf{\
depend on the ''angular'' type coordinates }$(\Phi ,\Psi ,\Theta )$\textbf{,
defined on the complex coordinates }$z$\textbf{\ and }$\overline{z}$\textbf{%
of the string world-sheet \ }
\begin{equation}
X^{\mu }\equiv X^{\mu }(\Phi (z,\overline{z}),\Psi (z,\overline{z}),\Theta
(z,\overline{z}))\text{ \ \ \ ,}  \tag{34}
\end{equation}

\textbf{where }$\mu =0,1,2,3$\textbf{, and both the }$ADS$\textbf{\
coordinates }$X^{\mu }$\textbf{\ and the variables }$(\Phi ,\Psi ,\Theta )$%
\textbf{\ are invariant under complex reparametrizations: }
\begin{equation}
X^{\mu }(z)\equiv X^{\mu }(w(z))\text{ \ ; \ \ \ }(\Phi (w(z)),\Psi
(w(z)),\Theta (w(z)))\equiv (\Phi (z),\Psi (z),\Theta (z))\text{ \ \ \ .}
\tag{35}
\end{equation}
From here by simple differentiation it can be obtained that
\begin{equation}
\frac{\partial \Phi (z)}{\partial z}-\frac{\partial \Phi (z)}{\partial w}%
\frac{\partial w}{\partial z}=0  \tag{36}
\end{equation}
(and of course, analogously for the other two derivatives $\frac{\partial
\Psi (z)}{\partial z}$ and $\frac{\partial \Theta (z)}{\partial z}$), but
with the important notice - only if the derivatives $\frac{\partial X^{\mu
}(z)}{\partial \Phi }$, $\frac{\partial X^{\mu }(z)}{\partial \Psi }$ and $%
\frac{\partial X^{\mu }(z)}{\partial \Theta }$ are \textbf{arbitrary}. It
will be proved in the paper \cite{17}, however, that if one specifies the
metric, for example in the form
\begin{equation}
ds^{2}=-cosh^{2}\rho dt^{2}+d\rho ^{2}+sinh^{2}\rho d\varphi ^{2}\text{ ,}
\tag{37}
\end{equation}
then the derivatives of $\ $the global $ADS$ coordinates $X^{\mu }$ for $\mu
=3$ are no longer independent.

2. The derivatives $\frac{\partial \Phi (z)}{\partial z}$, $\frac{\partial
\Psi (z)}{\partial z}$ and $\frac{\partial \Theta (z)}{\partial z}$ are
found by making use of the found algebraic relation (24), but at the point $%
v=0$, which means that the point $w$ is obtained from the point $z$ by
\textbf{an orbifold rotation at an angle }$\frac{\pi }{2}$\textbf{, i. e. }$%
w=\varepsilon e^{\frac{i\pi }{2}}z$ ($\varepsilon =\pm 1$). Consequently the
conformal generators $J_{R}^{3}(w)$, $J_{R}^{+}(w)$ and $J_{R}^{-}(w)$ are
obtained to be also orbifold rotated around the point $z$, i.e. $%
J_{R}^{3}(w)=-i\varepsilon J_{R}^{3}(z)$; $J_{R}^{+}(w)=-i\varepsilon
J_{R}^{+}(z)$ and \ $J_{R}^{-}(w)=-i\varepsilon J_{R}^{-}(z)$.

Further from the system of equations (4) for the conformal currents in the
OPE relations for the $SL(2,R)$ case, one can obtain the following system of
three first - order nonlinear differential equations for the variables $\Phi
,\Psi $ and $\Theta $:
\begin{equation}
\frac{\partial \Theta }{\partial z}=T(\Theta ,\Phi ,\Psi )\text{ ; \ \ \ \ }%
\frac{\partial \Phi }{\partial z}=Q(\Theta ,\Phi ,\Psi )T(\Theta ,\Phi ,\Psi
)\text{\ \ \ \ \ \ \ ; }  \tag{38}
\end{equation}
\begin{equation}
\frac{\partial \Psi }{\partial z}=P(\Theta ,\Phi ,\Psi )T(\Theta ,\Phi ,\Psi
)\text{ \ \ .}  \tag{39}
\end{equation}
Each function $\Phi ,\Psi $ and $\Theta $ is assumed to have \ a real and an
imaginary part ($\Phi (z)\equiv \Phi _{1}(z)+i\Phi _{2}(z)$ and etc.), so
after separating the real and the imaginary parts in each of the equations
and after combining (but not solving) the obtained equations, one can derive
the following simple consistency condition :
\begin{equation}
tanh2\Phi _{1}=\pm \frac{k}{L^{2}}\text{ \ \ .}  \tag{40}
\end{equation}
where $L$ is the DeSitter radius.

Now let $\Pi (\rho )$ is the \textbf{angle of parallelism in Lobachevsky
geometry}, which represents the angle between the perpendicular (through a
given point) towards a given line $l_{1}$ and the the parallel to $l_{1}$
line, drawn again through this point. After making the identifications
\begin{equation}
\Phi _{1}\equiv \frac{\Pi (\rho )}{4}\text{ \ \ \ \ \ \ ; \ \ \ \ \ \ }exp(-%
\frac{\rho }{r})=-i\varepsilon \frac{k}{L^{2}}\text{ \ \ \ ,}  \tag{41}
\end{equation}
the \textbf{Lobachevsky formulae for the angle of parallelism} is
recovered:\
\begin{equation}
\Pi (\rho )=2arctg(exp(-\frac{\rho }{r}))\text{ \ \ \ \ ,}  \tag{42}
\end{equation}
where $\rho $ is the hyperbolic distance $\rho =2\int\limits_{0}^{r}\frac{dx%
}{1-x^{2}}=2arctanhr$ $\ $(expressible also through the anharmonic
relation), and the simple formulae $tanh(ix)=itgx$ has been used.

\section{Inversion Formulaes from Integral Geometry - Application to the
Kac- Moody Algebra}

\bigskip The main purpose in this section will be to \textbf{apply the
inversion formulae (27)\ to the integral expression, which will be obtained
below from the Kac-Moody algebra (25). }In fact, this is the first step
towards constructing the integro - geometric approach for treating
simultaneously all the Kac - Moody and Virasoro algebras, based on applying
various methods from integral geometry. Some details will also be given in
\cite{17}.

The central idea here will be to reduce the double integration (in the L. H.
S.) in the integral representation of the commutator $\left[
J_{n}^{a},J_{m}^{b}\right] $ of \ the Kac-Moody algebra (25) to a single
integration by introducing a delta function $\delta (p(\zeta _{1},\zeta
_{2},z))$, defined on the already found $CP3$ hypersurface (24). Omitting
the intermidiate calculations, we get the following integral expression for
the Kac - Moody algebra (25):
\begin{equation}
\oint d\zeta _{2}\widetilde{G}(\zeta _{2},z)=\widetilde{F}(\zeta _{2},z)%
\text{ \ \ ,}  \tag{44}
\end{equation}
where $\widetilde{F}(\zeta _{2},z)$ and $\widetilde{G}(\zeta _{2},z)$ are
the expressions:
\begin{equation}
\widetilde{F}(\zeta _{2},z)\equiv \frac{1}{2}kn\delta ^{ab}\delta _{n+m,0}-%
\frac{f^{abc}}{2}J^{c}(\zeta _{2})(\zeta _{2}-z)^{n+m+3}t^{c}e^{-i\pi \left[
2(n+m)+3\right] }\text{ \ ,}  \tag{45}
\end{equation}
\begin{equation*}
\widetilde{G}(\zeta _{2},z)\equiv \frac{1}{2}J^{a}(\zeta _{1}=z-i(\zeta
_{2}-z))J^{b}(\zeta _{2})t^{a}t^{b}(\zeta _{2}-z)^{m+n+3}\times
\end{equation*}
\begin{equation}
\times \left[ \varepsilon ^{m+1}e^{-\frac{i(m+3)\pi }{2}}-\varepsilon
^{n+1}e^{-\frac{i(n+3)\pi }{2}}\right] \text{ \ \ \ .}  \tag{46}
\end{equation}
In order to find the ''inverse'' formulae for the integral representation
(44), or in other words - an expression for $\widetilde{G}(\zeta _{2},z)$ as
an integral, depending on the function $\widetilde{F}(\zeta _{2},z)$, we
shall search for a representation of this function in the form: \
\begin{equation}
\widetilde{F}(\zeta _{2},z)=\int P(X)\delta (\left[ X,\eta \right] -1)dX%
\text{ \ \ \ .}  \tag{47}
\end{equation}
\bigskip\ Now it remains to find the function $P(X)$ by making use of the
integral inversion formulae (27):
\begin{equation}
P(X)=-\frac{1}{8\pi ^{2}}\widetilde{F}(\zeta _{2},z)K(X)=-\frac{1}{8\pi ^{2}}%
\widetilde{F}(\zeta _{2},z)\int \delta ^{^{\prime \prime }}(\left[ X,\eta %
\right] -1)d\eta \text{ \ \ ,}  \tag{48}
\end{equation}
then to substitute it into the integral representation formulae (47)\ \ and
finally, to compare the both sides of (44). The peculiar moment here is that
in the L. H. S. of \ (44) the integration is over the $\zeta _{2}$
variables, while in the R. H.\ S. it is over the variables $\eta $ and $%
X=(X^{0},X^{1},X^{2},X^{3})$. And since $X^{0\text{ }}$can be expressed
through the hyperboloid equation, it remains to integrate over $%
(X^{1},X^{2},X^{3})$. In order to be able to compare both sides of (44), we
shall pass from an integration over \ $(X^{1},X^{2},X^{3})$ to an
integration over $(\zeta _{2},z)$ by applying the formulae for the
transformation Jacobian:
\begin{equation}
dX^{1}\wedge dX^{2}\wedge dX^{3}\equiv \frac{\partial (X^{i_{1}},X^{i_{k}})}{%
\partial (\zeta _{2},z)}=\sum\limits_{1\leq i_{1}<i_{k}<3}det\left(
\begin{array}{cc}
\frac{\partial X^{i_{1}}}{\partial \zeta _{2}} & \frac{\partial X^{i_{1}}}{%
\partial z} \\
\frac{\partial X^{i_{k}}}{\partial \zeta _{2}} & \frac{\partial X^{i_{k}}}{%
\partial z}
\end{array}
\right) d\zeta _{2}\wedge dz\text{ \ \ ,}  \tag{49}
\end{equation}
where $\frac{\partial X^{i_{1}}}{\partial \zeta _{2}}$ simply denotes $\frac{%
\partial X^{i_{1}}}{\partial z}$, taken at the point $z=\zeta _{2}$. The
inversion formulae for $\widetilde{G}(\zeta _{2},z)$ in (44) is obtained in
the form
\begin{equation}
\widetilde{G}(\zeta _{2},z)=-\frac{1}{8\pi ^{2}}\int d\eta dz\text{ }%
\widetilde{F}(\zeta _{2},z)\text{ }\delta (\left[ X,\eta \right] -1)\delta
^{^{\prime \prime }}(\left[ X,\eta \right] -1)\frac{\partial
(X^{i_{1}},X^{i_{k}})}{\partial (\zeta _{2},z)}N(X)  \tag{50}
\end{equation}
of a complicated nonlinear differential equation in respect to the variables
$X$. The function $N(X)$ comes from the variable change $%
(X^{0},X^{1},X^{2},X^{3})\rightarrow (X^{1},X^{2},X^{3})$, which is
calculated by means of the generalization of the Jacobian (49) for a higher
dimensional case.

\textbf{Acknowledgements.} The author is greateful to Prof. N. A. Chernikov
and Dr. N. S. Shavokhina for the kind invitation to make a report at the
International Seminar on Lobachevsky Geometry (26-30 February, 2004, BLTP,
JINR, Dubna), dedicated to the 75th Anniversary \ of \ Prof. N. A.
Chernikov. The author is grateful also to Prof. I. Todorov, to Prof. S.
Manoff and to Dr. N. Nikolov (INRNE, BAS, Sofia, Bulgaria) for useful
conversations on CFT and integral geometry and for their interest towards
this research. This paper was partially supported also by a Shoumen
University grant \ 005/2002.



\end{document}